\documentclass[
    ,final            
  ]
  {aipproc}

\layoutstyle{8x11single}

\usepackage{amssymb}

\begin{document}

\title{Measurement of Neutrino-Nucleon Neutral Current Elastic Scattering in MiniBooNE}


\author{Denis Perevalov$^*$ and Rex Tayloe$^\dagger$ for the MiniBooNE collaboration}{address={$^*$University of Alabama, Tuscaloosa AL\\$^\dagger$Indiana University, Bloomington IN}}

\begin{abstract}
Using a high-statistics sample of neutral current elastic neutrino interactions,
MiniBooNE measured the flux-averaged neutral current elastic differential cross-section on mineral oil ($CH_2$).
Using the latter, a $\chi^2$ test of MC with different values of the axial vector mass has been performed.
Also, a possibility of using a sample of neutral current elastic proton-enriched events above Cherenkov threshold
to measure the ratio $\nu p\rightarrow \nu p /\nu N\rightarrow \nu N$ is discussed.
This ratio is sensitive to the strange quark contribution to the nucleon spin, $\Delta s$.
\end{abstract}
\maketitle

\section{Neutral Current Elastic Interactions in MiniBooNE}
Neutrino-nucleus neutral current elastic (NCE) scattering, $\nu N \rightarrow \nu N$ ($N=p,n$), is a fundamental probe of the nucleon.
This interaction, in principle, is sensitive to the isoscalar part of the weak neutral current axial form factor, especially when it is possible to
separate protons from neutrons.
In MiniBooNE, NCE events account for about 18\% of all neutrino interactions.
It seems virtually impossible to measure these interactions in a Cherenkov detector, because most of the protons have an energy below
Cherenkov threshold.
Also, there is not enough light for tagging the neutron capture to separate neutrons from protons.
However, the use of mineral oil as a medium in the MiniBooNE detector~\cite{MB_detec} allows
for a measurement of these interactions even in a region below Cherenkov threshold, because of the presence of fluors in the mineral oil.
Hence, charged particles going through the medium produce scintillation and fluorescence light in addition to any Cherenkov light.

There was a preliminary measurement of the MiniBooNE  NCE differential cross-section~\cite{Cox_nuint}, where
a reduced MiniBooNE data set with $6.57\times10^{19}$ protons on target (POT) was used, at the time restricted by the oscillation blindness requirement.
Since then, MiniBooNE has improved the analysis, as well as used the full data set with $6.46\times10^{20}$ POT for the latest NCE results. 
The NCE flux-averaged differential cross-section has been measured and is reported.
It has also been used for the $\chi^2$ tests of MC with different values of the axial vector mass, $M_A$.

NCE scattering may occur off either protons or neutrons.
In the detector, such an interaction is observed via a recoil nucleon.
As opposed to protons, neutrons are neutral particles and do not produce light by themselves. 
They are normally observed in the detector via a charge exchange process $np\rightarrow np$ that follows the initial neutrino interaction.
For this reason, in MiniBooNE it is impossible to distinguish between protons and neutrons for protons below 
Cherenkov threshold in mineral oil (i.e., for proton kinetic energy $T\leqslant 350$ MeV).

Because of the impossibility of measuring the outgoing neutrino in such interactions,
a stationary target nucleon is assumed in the momentum transfered squared ($Q^2$) reconstruction.
By measuring the outgoing nucleon kinematics, the reconstructed $Q^2$ may be calculated as:
\begin{equation}
Q^2 = 2m_N T,
\label{eq:q2}
\end{equation}
where $m_N$ is the nucleon mass and $T$ is the reconstructed nucleon kinetic energy.

The \textit{signal} consists of 3 different processes:
$\nu N\rightarrow \nu N$ scattering off bound protons and neutrons inside carbon atoms and off hydrogen atoms.
Thus, MiniBooNE measures the NCE flux-averaged differential cross-section, which also averages over the signal processes.
The weight for each signal process in the total NCE cross-section is given in Eq.\eqref{eq:NCE_xs_expression}.

Neutrino interactions within the detector are simulated with the v3 Nuance
event generator~\cite{nuance}, where the relativistic Fermi gas model of Smith and Moniz~\cite{SmithMoniz} is used to describe NCE interactions.
For resonant pion production the Rein and Sehgal model~\cite{ReinSehgal} is used. 
In the few GeV range, such interactions are dominated by the $\Delta(1232)$ resonance, although, contributions from higher mass resonances are also included.
The values of the parameters assumed in the neutrino cross-section model and their uncertainty are shown in Table~\ref{tab:xs_parameters}.

\begin{table}
\begin{tabular}{llll}
\hline
    \tablehead{1}{l}{b}{Parameter}
  & \tablehead{1}{l}{b}{Value}
  & \tablehead{1}{l}{b}{Uncertainty}
  & \tablehead{1}{l}{b}{Description}\\
\hline
M\_A QE                 & $1.2341$ GeV & $0.077$ GeV    &$M_A$ for QE events on carbon          \\ 
E\_B                    & $34.0$ MeV  & $9.0$ MeV        &Binding energy for carbon            \\ 
p\_F                    & $220.0$ MeV & $30.0$ MeV      &Fermi momentum for carbon            \\ 
dels                    & $0.0$    & $0.1$             &$\Delta s$, the axial vector isoscalar term\\ 
M\_A 1pi                & $1.1$ GeV & $0.27$ GeV        &$M_A$ for CC and NC 1pi events         \\ 
M\_A Npi                & $1.3$ GeV& $0.52$ GeV        &$M_A$ for CC and NC multipion events   \\ 
coh                     & $1.302$  & $0.14$          &Scale factor for NC coherent pi0 events\\ 
delrad                  & $1.00$   & $0.12$           &Scale factor for NC and CC delta radiative events\\ 
dis                     & $1.00$   & $ 0.25$          &Scale factor for deep inelastic scattering events\\ 
EloSF                   & $1.0220$ & $ 0.0205$      &Pauli blocking $\kappa$              \\ 
M\_A Coh                &  $1.030$ GeV & $0.275$ GeV    &$M_A$ for CC1pi coherent events (not NCpi0 coherent)\\  
Res Pi0                 &  $1.00$  & $0.14$          &Scale factor for NC resonant pi0 events \\  
M\_A QE H               & $1.13$ GeV & $0.10$ GeV      &$M_A$ for QE events on Hydrogen     \\ 
\hline
\end{tabular}
\caption{Cross-section parameters and their 1-$\sigma$ uncertainties used in the MiniBooNE MC.}
\label{tab:xs_parameters}
\end{table}

Intranuclear final state interactions (FSI) inside ${}^{12}C$ are also interpreted in Nuance, where the scattered hadrons
are propagated through the nucleus, which is simulated based on models of nuclear density and Fermi momentum.
At each step, the MC determines whether the hadron interacts with the target nucleus
and if so, generates the interaction and the resulting new final state particles.
Particle propagation in the detector is a GEANT3-based~\cite{GEANT3} MC 
with GCALOR~\cite{gcalor} hadronic interactions simulating the detector response to particles produced in the neutrino interactions.
Due to the FSI, a NCE event may have more than 1 final state particle.
Also, a NC pion event may have no pion in the final state as the pion can get absorbed within the carbon nucleus.
According to Nuance, for NCE scattering on a target nucleon inside ${}^{12}C$, the probability of having only 1 outgoing nucleon is $\sim74\%$.
Also, the probability that the outgoing pion experiences either absorption or charge exchange is $\sim 20\%$.

Neutrino events that happen outside of the detector in the dirt or in the detector material (referred to as "dirt" background henceforth)
are simulated the same way as in-tank events but with a cross-section reweighed according to the density of the material relative to the density of the mineral oil.

\section{Event Reconstruction and Selection. Dirt Background Constraint.}
One of the main analysis improvements is the NCE event reconstruction, which assumes that an event is a single proton. 
The fitter looks for a set of parameters (position, time, direction and energy) 
which maximizes the event likelihood using the charge and time information from all photomultiplier tubes (PMTs).
The event is assumed to be point-like with the Cherenkov and scintillation profiles for protons measured from the MC.
The position resolution is $\sim0.75$ m, the energy resolution $\sim 25\%$, and for protons above Cherenkov threshold the angular resolution is $\sim 10^o$.
The reconstructed nucleon kinetic energy, $T$, is proportional to the total charge on all PMTs, 
and thus it is a measure of the sum of the kinetic energies of all final state nucleons that are produced in the interaction.

\begin{figure}
\includegraphics[width=0.5\textwidth]{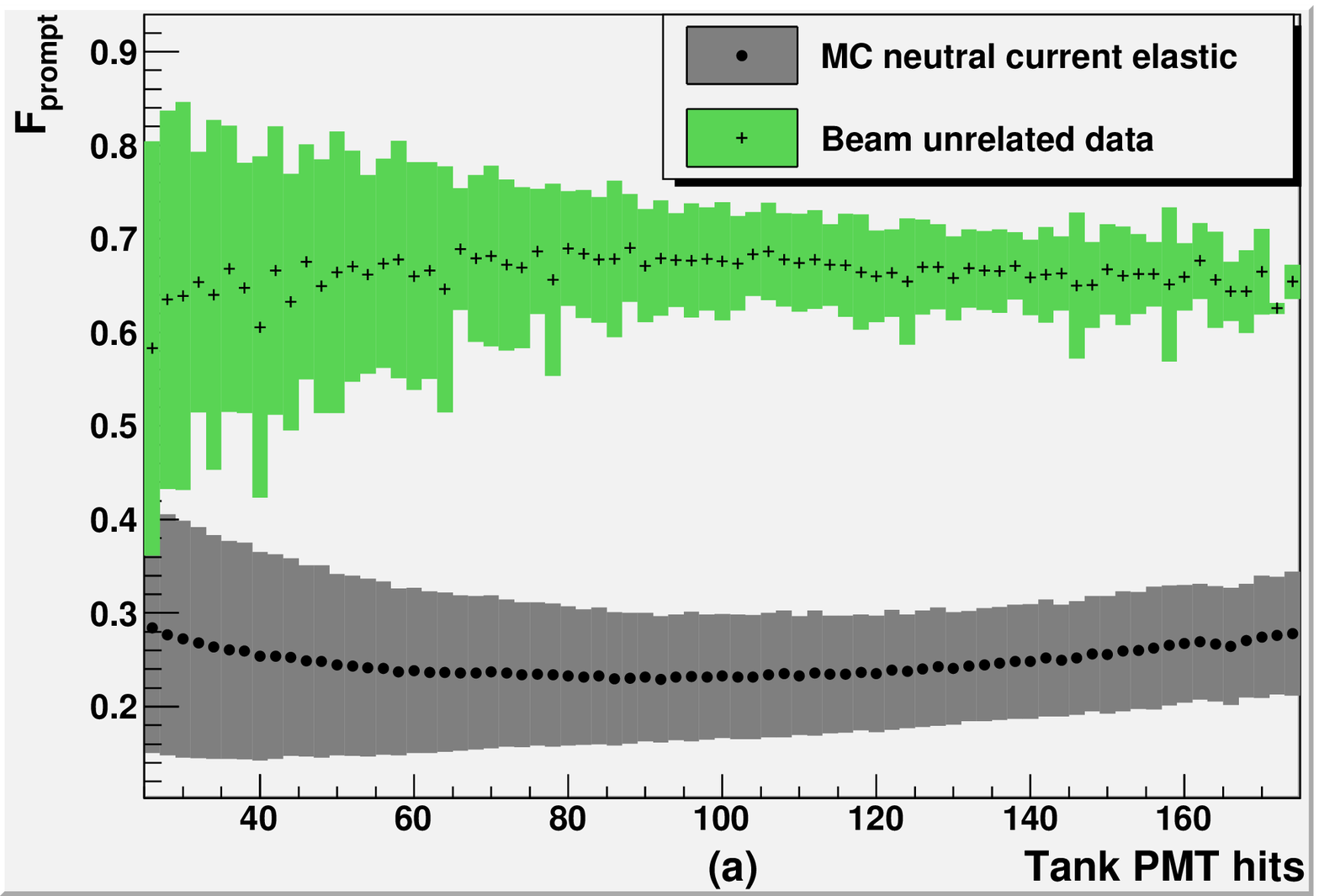}
\includegraphics[width=0.5\textwidth]{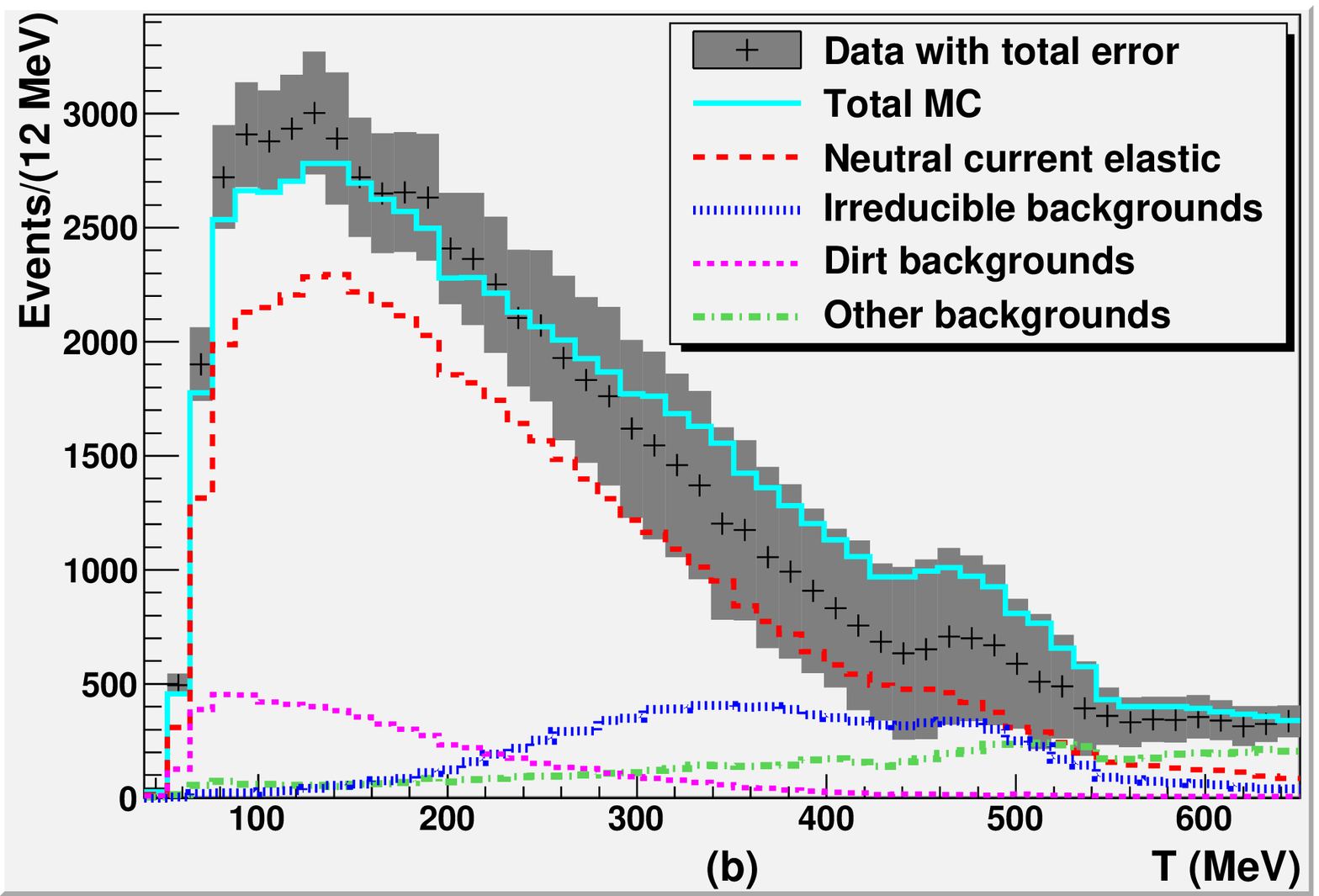}
\caption{(a): Prompt light fraction (hits with $-5\ ns<t_{corr}<5\ ns$) for the beam unrelated data and the NCE MC under electron hypothesis reconstruction.
(b): Reconstructed nucleon kinetic energy spectra for the data and different MC channels after the NCE selection cuts applied, but using a uniform fiducial volume cut of $R<4.2$ m.
Distributions are absolutely (POT) normalized.}
\label{fig:erec}
\end{figure}

The following set of cuts are applied to the full MiniBooNE data set
to select the NCE sample:
 (1) only $1\ subevent$ to ensure no decaying particles ($\mu$ or $\pi$ decay), 
 (2) veto PMTs $< 6$ to remove events exiting or entering the detector, 
 (3) tank PMTs $>25$ to use only reconstructable events,
 (4) beam time window cut, 
 (5) a cut on the reconstructed proton energy $T<650$ MeV (above which the NCE signal to background ratio decreases significantly), 
 (6) a cut on the time log-likelihood ratio between proton and electron event hypotheses $\ln(\mathcal{L}_{e}/\mathcal{L}_{p})<0.42$ to eliminate
beam-unrelated electrons from cosmic ray muon decays
(From Fig.\ref{fig:erec}(a) one can see that electrons have more prompt light fraction than protons), 
 (7) a fiducial volume cut, $R<4.2$ m if $T<200$ MeV and $R<5$ m otherwise. The reason to have a tighter fiducial volume at low energies is to reduce dirt backgrounds.

A total of 94,531 events pass the NCE cuts in the sample corresponding to $6.46\times 10^{20}$ POT.
The efficiency of the cuts is estimated to be 26\% (a large part of the reduction is due to the fiducial volume cut).
The fraction of NCE events in the sample is 65\%.
The remaining 35\% of events in the sample are backgrounds:
15\% irreducible backgrounds,
10\% dirt events,
and 10\% other backgrounds (of which only 0.5\% are beam unrelated).
The reconstructed nucleon kinetic energy spectrum for the selected NCE event sample is shown in Fig.\ref{fig:erec}(b).

Irreducible backgrounds are those neutral current $\pi$ production channels that have no $\pi$ in the final state, as the $\pi$ is absorbed in the nucleus through a FSI process.
In this case, the final state particles for these events are just nucleons, identical to the final state particles produced in NCE events and hence are "irreducible".
These backgrounds are important at intermediate energies, $200$ MeV$\lesssim T \lesssim 500$ MeV.

The dirt background is an important contribution to the NCE data sample at low energies, mostly below 200 MeV.
Normally, this background is due to nucleons (mainly neutrons), which are produced in neutrino interactions outside of the detector,
penetrating into the detector without firing enough veto PMTs.
Dirt events are difficult to simulate in the MC because they happen outside of the detector in the soil, whose properties have not been studied in detail.
It is also hard to correctly describe the particle propagation through a set of different materials (through the soil, detector parts and into the detector media). 
However, it is possible to directly measure these backgrounds in the data, as they have different kinematic observables than in-tank events,
such as the reconstructed radius $R$ (dirt events are mostly in outer regions),
the reconstructed $Z$
(dirt events are mostly upstream with $Z<0$) and the reconstructed energy (dirt events are at $T\lesssim 200$ MeV).
%
The dirt energy spectrum is measured by fitting the MC in-tank and dirt templates to the data in $Z$, $R$, and energy distributions
for the dirt-enriched event samples.
All of these measurements agree with each other within 10\%.
The measured number of dirt events is determined to be $\sim30\%$ lower than the original MC prediction.

\section{Final Neutral Current Elastic Differential Cross-Section Result.}
The obtained reconstructed energy spectrum is then corrected for limited detector resolution and detector efficiency effects that distort the original spectrum.
This is the so-called unfolding problem~\cite{Cowan_text}, for which a method based on Bayes' theorem~\cite{Agostini} is used.
This method gives a well-behaved but biased solution which depends on the original MC energy spectrum. 
The error due to the bias is estimated by iterating the unfolding procedure, where the new MC energy spectrum is replaced by the unfolded energy spectrum.

\begin{figure}
\includegraphics[height=2.6in, width=3.5in]{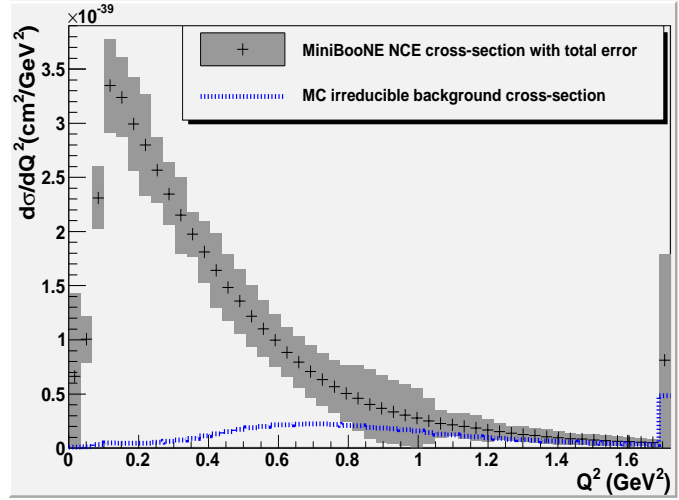}
\caption{MiniBooNE $\nu N\rightarrow \nu N$ flux-averaged differential cross-section on $CH_2$ along with the predicted irreducible background cross-section, that has been subtracted out.
}
\label{fig:xs}
\end{figure}

Possible systematic uncertainties and their contribution to the total error have been studied and are shown in Table~\ref{tab:errors}.
With this, the NCE flux-averaged differential cross-section is shown in Fig.\ref{fig:xs} as a function of $Q^2$.
The predicted distribution of the irreducible backgrounds, which have been subtracted along with the rest of the backgrounds, is also shown in the figure.

\begin{table}
\begin{tabular}{lc}
\hline
    \tablehead{1}{l}{b}{Error}
  & \tablehead{1}{l}{b}{Value,\%}\\
\hline
    Statistical             \hspace{8cm}             & 3.1         \\
    Discriminator threshold                          & 1.1         \\
    QT PMT response                                  & 2.0         \\
    Dirt                                             & 1.4         \\
    POT                                              & 1.7         \\
    $\pi^0$ yield                                    & 0.1         \\
    Hadronic                                         & 0.2         \\
    Cross-section                                    & 3.4         \\
    Horn uncertainties                               & 4.7         \\
    $K_0$  production                                & 0.1         \\
    $K^+$  production                                & 0.5         \\
    $\pi^-$  production                              & 0.3         \\
    $\pi^+$  production                              & 4.1         \\
    Optical model                                    & 12.3        \\
    Unfolding                                        & 7.5         \\ \hline
    Total                                            & 18.9        \\
\hline
\end{tabular}
\caption{The individual error contributions to the total error for the reconstructed energy spectrum and true energy spectrum (or the NCEL cross-section error).}
\label{tab:errors}
\end{table}

\begin{figure}
\includegraphics[width=0.5\textwidth]{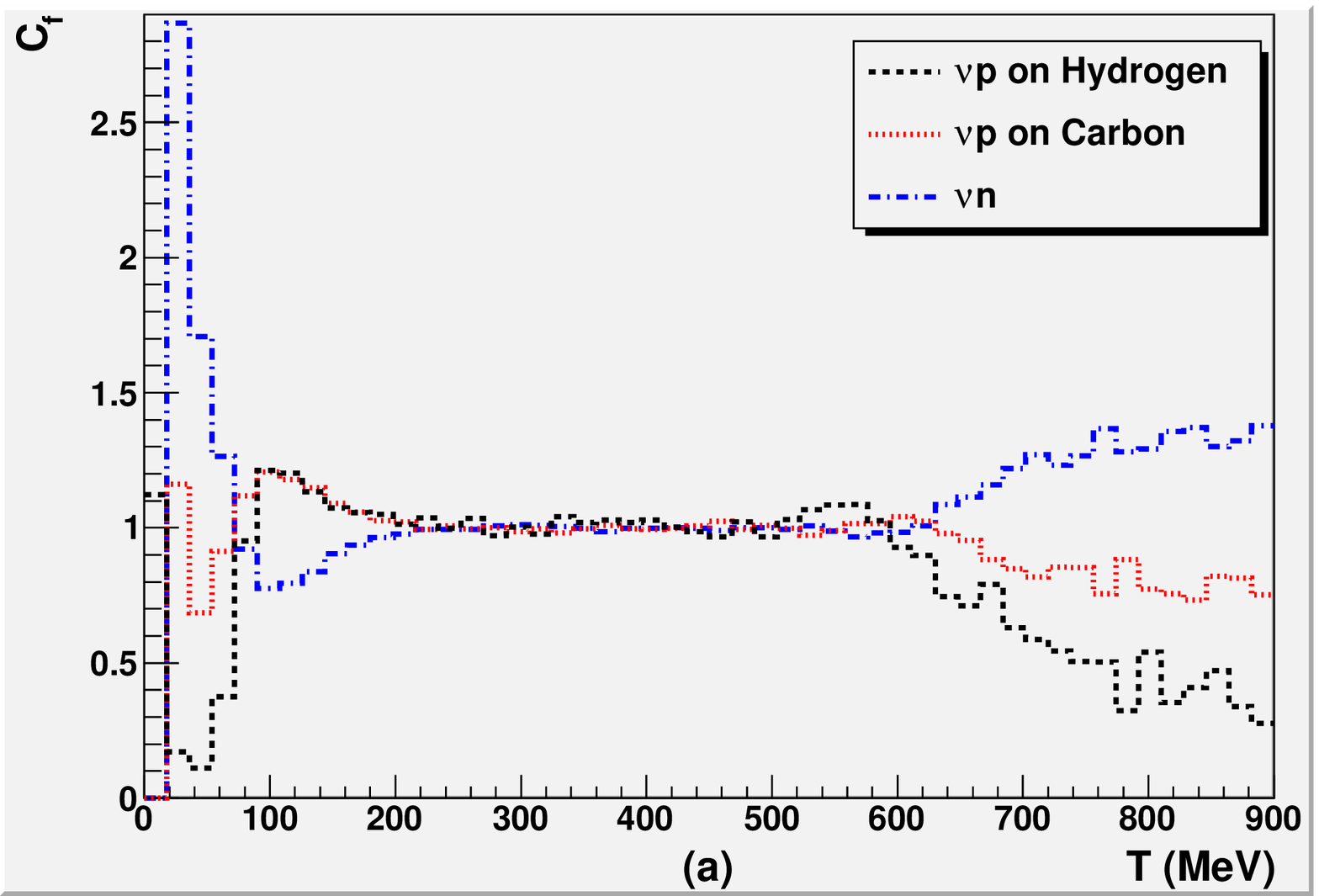}
\includegraphics[width=0.5\textwidth]{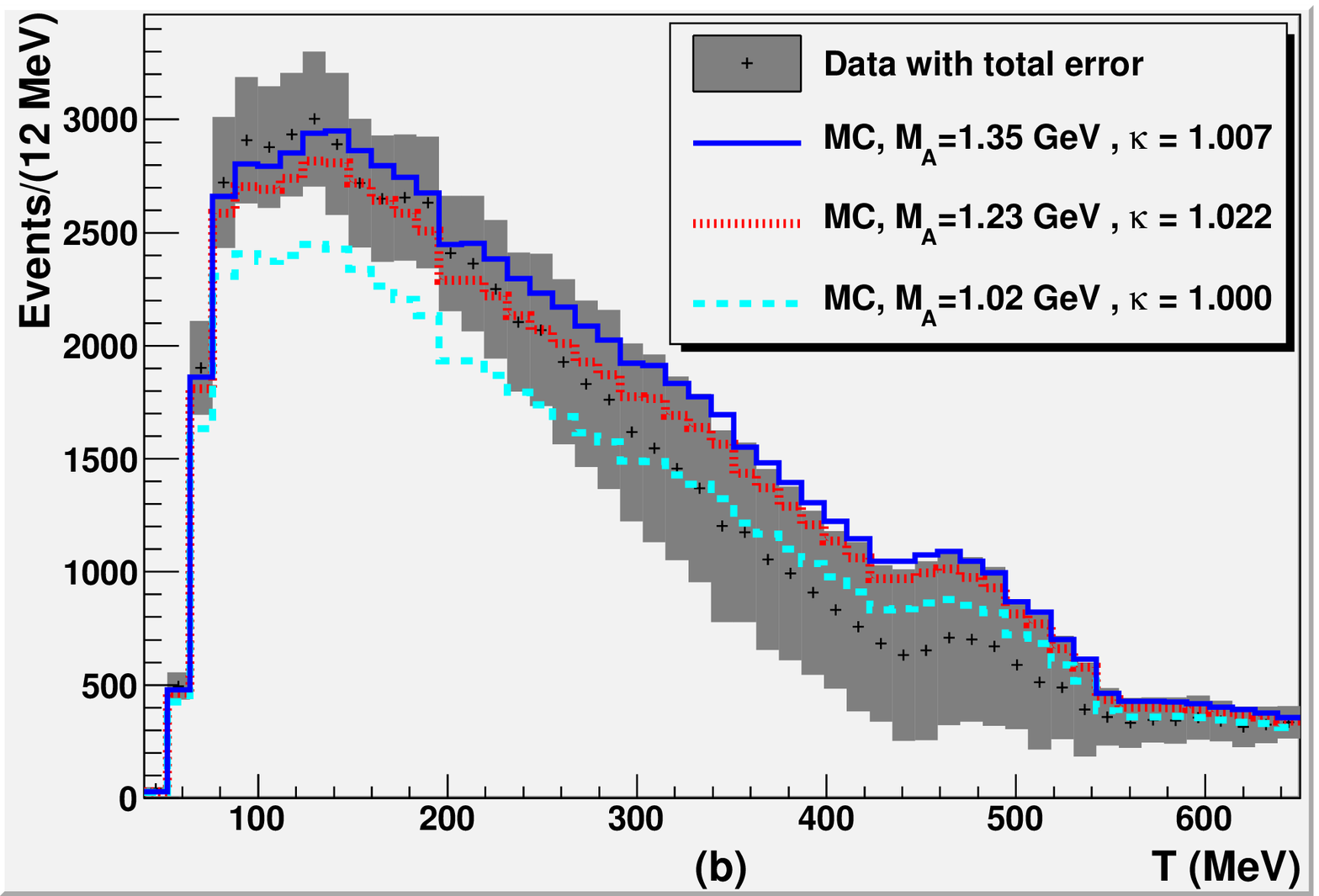}
\caption{
(a): Efficiency corrections, $C_f$, as a function of the nucleon kinetic energy for different NCE processes.
(b): $\chi^2$-tests on the NCE reconstructed energy for $M_A$ values of 1.35, 1.23, and 1.02 GeV. 
The $\chi^2$ values are 27.1, 29.2 and 41.3 for 49 DOF, respectively. The distributions are absolutely normalized.}
\label{fig:MA_results}
\end{figure}

The average chemical formula for mineral oil is CH$_2$. Thus, the $\nu N\rightarrow \nu N$ cross-section 
in terms of the NCE cross-section on different types of bound and free nucleons is expressed as:
\vspace{-0.0in}
\begin{equation}
   \frac{d\sigma_{NCE}}{dQ^2} = 
   \frac1{7}C_{\nu p\rightarrow \nu p,H}(Q^2)\frac{d\sigma_{\nu p\rightarrow \nu p,H}}{dQ^2} 
  +\frac3{7}C_{\nu p\rightarrow \nu p,C}(Q^2)\frac{d\sigma_{\nu p\rightarrow \nu p,C}}{dQ^2}  
  +\frac3{7}C_{\nu n\rightarrow \nu n,C}(Q^2)\frac{d\sigma_{\nu n\rightarrow \nu n,C}}{dQ^2},  
\label{eq:NCE_xs_expression}
\end{equation}
where the first term is the NCE cross-section on free protons taken per free proton, etc.
The efficiency correction functions $C_{f}$ are due to different efficiencies for the signal processes.
They are defined as the ratio of the efficiency for a particular type of signal events to the average efficiency for all signal events.
In principle, one needs to know the efficiency corrections $C_{\nu p,H}$, $C_{\nu p,C}$ and $C_{\nu n,C}$, which are shown in Fig.\ref{fig:MA_results}(a).
But for some quick calculations one can roughly assume $C_{\nu p,H}=C_{\nu p,C}=C_{\nu n,C}=1$.

The NCE cross-section is quite sensitive to the axial form factor. In fact, at low $Q^2$, $d\sigma /dQ^2\sim G_A^2(Q^2)$.
The axial form factor is usually taken in the dipole approximation:
\vspace{-0.0in}
\begin{equation}
   G_A(Q^2) = 0.5(\mp g_A+\Delta s)/(1+Q^2/M_A^2)^2 
\label{g_a_expression}
\end{equation}
where $g_A=1.2671$,
$M_A$ is the axial vector mass,
the $+$ sign is for protons, and the $-$ sign is for neutrons.

Inserting Eq.(\ref{g_a_expression}) for protons and neutrons into Eq.(\ref{eq:NCE_xs_expression}),
one can see that the MiniBooNE $\nu N\rightarrow \nu N$ NCE cross-section is not sensitive to $\Delta s$,
as the linear term in $\Delta s$ nearly cancels, while the quadratic term in $\Delta s$ remains, but it is small if $|\Delta s|\ll g_A$.
However, one can still use these data to probe $M_A$.

There seems to be some recent interest in the value of $M_A$.
Before MiniBooNE, $M_A$ has been measured by mostly bubble chamber deuterium-based neutrino experiments, which found $M_A=1.02$ GeV~\cite{MA_ave}.
However, recent measurements on carbon, oxygen and iron targets by MiniBooNE\cite{MB_ccqe_nuint09}, K2K~\cite{K2K_ccqe,K2K_ccqe2} and MINOS~\cite{Minos_nuint09}
have determined it to have $\sim 20-30\%$ larger values, using the CCQE channel.
In the MiniBooNE MC it is possible to generate the reconstructed energy spectra for NCE samples that correspond to any value of $M_A$. 
Here we test how different values of $M_A$ agree with the data assuming absolutely normalized distributions of nucleon energy.
To calculate the $\chi^2$ between data and MC, the full error matrix has been used which takes into account
the variations of MC energy spectrum due to various MC uncertainties (such as flux, cross-section, beam etc).
The reconstructed energy spectrum for the data and MC for $M_A$ and $\kappa$ values that we are interested in are shown in Fig.\ref{fig:MA_results}(b).
The higher values of $M_A$ seem to describe the MiniBooNE NCE data better.

\section{High Energy Proton-Enriched Sample. $\Delta s$ Sensitivity.}
As mentioned before, the $\Delta s$ sensitivity of the NCE sample comes down to the possibility of distinguishing proton from neutron events.
In particular, a ratio of $\nu p\rightarrow \nu p$ to $\nu n\rightarrow \nu n$ is most sensitive to $\Delta s$ in addition to reducing the systematic error.

In MiniBooNE, it is virtually impossible to distinguish protons from neutrons, because 
neutrons are usually only registered after a strong interaction with a proton.
However, it is possible to select a special class of NCE protons, namely those with only 1 proton in the final state
and energies above the proton Cherenkov threshold. These events can be used for a $\Delta s$ measurement 
relative to the total measured $\nu N\rightarrow \nu N$ cross-section.

It should be noted that events with multiple nucleons in the final state (NCE neutrons and irreducible backgrounds) have multiple protons produced in the reaction.
At kinetic energies above 350 MeV, single proton events produce on average a higher Cherenkov light fraction than multiple proton events.
In addition, these classes differ in the kinematics of the outgoing nucleon, with single proton events being more forward-going.

In order to reduce systematic uncertainties, a ratio $\nu p\rightarrow \nu p /\nu N\rightarrow \nu N$ as a function
of reconstructed nucleon kinetic energy from 350 MeV to 800 MeV is used.
The denominator of the ratio are events with NCE selection cuts used in the cross-section
measurement, but the energy cut (5) has been replaced with $350$ MeV$<T<800$ MeV,
and an additional cut is implemented, $\ln(\mathcal{L}_{\mu}/\mathcal{L}_{p})$
(a log likelihood ratio between muon and proton hypotheses) in order to reduce muon-like backgrounds that dominate at high visible energy region.
With these cuts, the $\nu N\rightarrow \nu N$ data sample has 24,004 events with predicted channel fractions:
45\% NCE, 26\% irreducible backgrounds, 3\% dirt backgrounds and 25\% other backgrounds.

\begin{figure}
\includegraphics[width=0.5\textwidth]{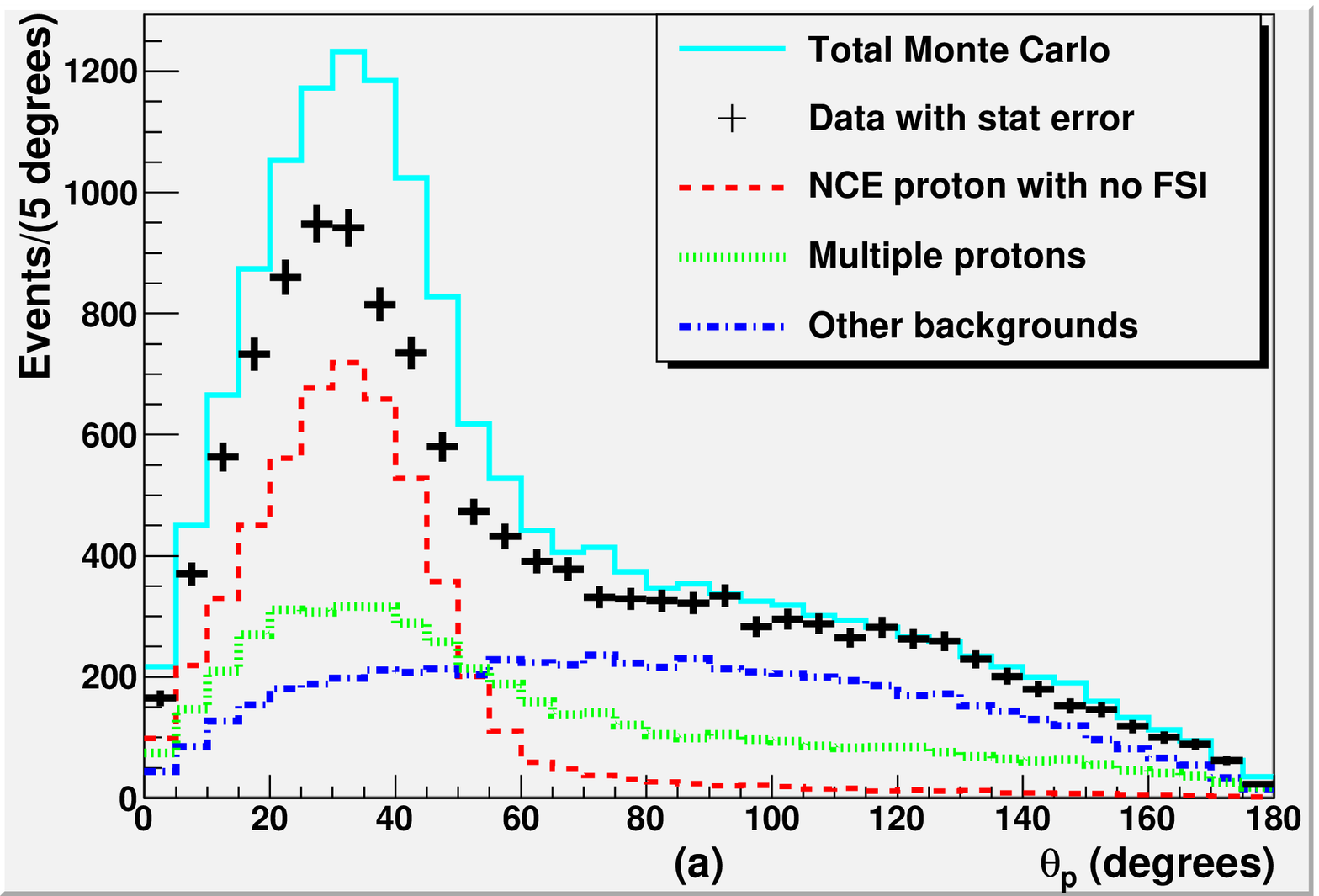}
\includegraphics[width=0.5\textwidth]{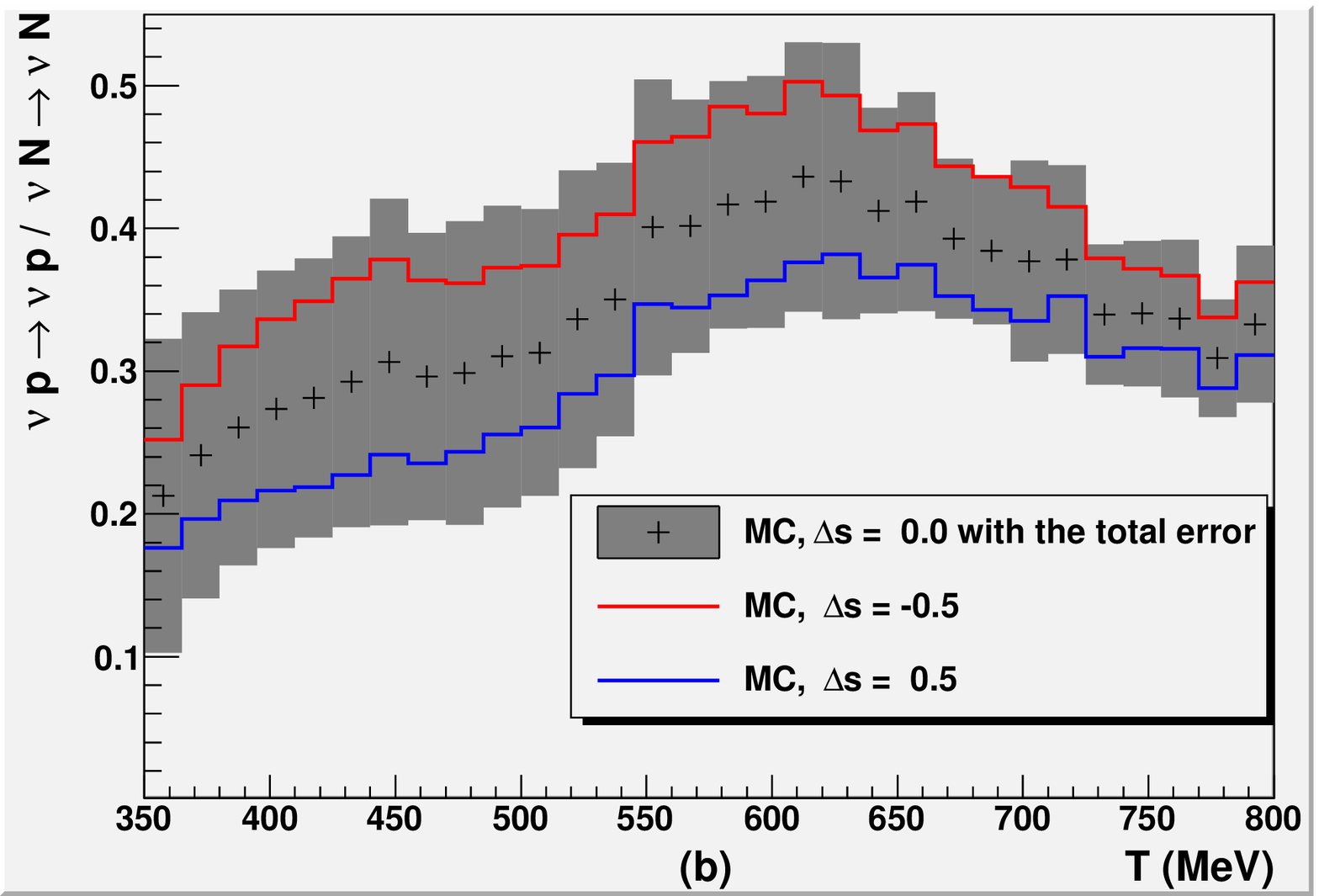}
\caption{
(a): $\theta_p$, the angle between the nucleon reconstructed direction and the beam direction for data and MC after NCE and $p/\mu$ cuts are applied.
All distributions are absolutely normalized.
(b): The ratio $\nu p\rightarrow \nu p /\nu N\rightarrow \nu N$ as a function of the reconstructed energy for MC with $\Delta s=0$ with total error.
Also MC with $\Delta s=-0.5 \mbox{ and }0.5$ are shown.
}
\label{fig:thetap}
\end{figure}

For the numerator, two more cuts are applied in addition to the ones used for the denominator.
The first cut is based on the prompt light fraction. 
It reduces multiple proton events, which have less Cherenkov light than single proton events. 
After this cut, looking at the $\theta_p$ distribution (the angle between the reconstructed nucleon direction and the beam direction), 
shown in Fig.\ref{fig:thetap}(a), one can see that most of the single proton events are forward going.
Thus, the last cut applied for the NCE p sample is $\theta_p<60^o$. The final $\nu p\rightarrow \nu p$ data sample has 7,616 events with predicted
channel fractions: 55\% NCE p, 10\% NCE n, 14\% irreducible backgrounds, 1\% dirt backgrounds, and 19\% other backgrounds.
The ratio of $\nu p\rightarrow \nu p$ to $\nu N\rightarrow \nu N$ for the MC with $\Delta s=0$ and the total uncertainties 
along with the MC with values of $\Delta s=-0.5 \mbox{ and }0.5$ is shown in Fig.\ref{fig:thetap}(b).

This will be the first attempt at a $\Delta s$ measurement using this ratio.
The systematic errors are quite large, mostly due to large uncertainties in the optical model of the mineral oil.
MiniBooNE has sensitivity to $\Delta s$ for proton energies above Cherenkov threshold for protons, where the contribution from irreducible backgrounds is significant.
In order to have a good sensitivity for $\Delta s$, future experiments need to have good proton/neutron identification (possibly through neutron capture tagging)
and extend down $T<200$ MeV region, where the contribution from irreducible backgrounds is less.

In summary, MiniBooNE has used a high-statistics sample of NCE interactions
to measure the total NCE cross-section.
Using absolutely normalized distributions of the reconstructed energy for the NCE sample,
$\chi^2$ tests for several $M_A$ and $\kappa$ values have been performed.
The MC with higher values of $M_A$, 1.23 GeV and 1.35 GeV gives a better $\chi^2$, than 1.02 GeV,
which is in agreement with the shape normalized fits of the CCQE channel obtained by recent experiments.
For energies above Cherenkov threshold, a sample of NCE proton-enriched protons was obtained. 
These events allow a measurement of the ratio of $\nu p\rightarrow \nu p$ to $\nu N\rightarrow \nu N$ and is sensitive to $\Delta s$.
The approximate sensitivity plot for the MiniBooNE $\Delta s$ extraction was shown.

\bibliographystyle{kickass}

\bibliography{nuint}

\IfFileExists{\jobname.bbl}{}
 {\typeout{}
  \typeout{******************************************}
  \typeout{** Please run "bibtex \jobname" to optain}
  \typeout{** the bibliography and then re-run LaTeX}
  \typeout{** twice to fix the references!}
  \typeout{******************************************}
  \typeout{}
 }

\end{document}